\documentclass[superscriptaddress,twocolumn]{revtex4}

\usepackage{graphicx}
\usepackage{times}
\usepackage{epsfig}
\usepackage{amsmath}
\usepackage{amsfonts}
\usepackage{amssymb}
\usepackage{url}
\usepackage{hyperref}
\newcommand{\be}{\begin{equation}}
\newcommand{\ee}{\end{equation}}
\newcommand{\bea}{\begin{eqnarray}}
\newcommand{\eea}{\end{eqnarray}}

\begin{document}

\pagestyle{plain}

\title{Light Higgs Mass Bound in SUSY Left-Right Models}

\author{Yue Zhang}
\affiliation{Center for High-Energy Physics and Institute of
Theoretical Physics, Peking University, Beijing 100871, China}
 \affiliation{Maryland Center for Fundamental Physics and Department of Physics, University of
Maryland, College Park, Maryland 20742, USA }
\author{Haipeng An}
 \affiliation{Maryland Center for Fundamental Physics and Department of Physics, University of
Maryland, College Park, Maryland 20742, USA }
\author{Xiangdong Ji}
 \affiliation{Maryland Center for Fundamental Physics and Department of Physics, University of
Maryland, College Park, Maryland 20742, USA } \affiliation{Center
for High-Energy Physics and Institute of Theoretical Physics, Peking
University, Beijing 100871, China}
\author{Rabindra N. Mohapatra}
 \affiliation{Maryland Center for Fundamental Physics and Department of Physics, University of
Maryland, College Park, Maryland 20742, USA }

\date{April, 2008}

\preprint{\vbox{\hbox{UMD-PP-08-00X}}}

\begin{abstract}

We show that in supersymmetric left-right models (SUSYLR), the upper
bound on the lightest neutral Higgs mass can be appreciably higher than that in
minimal supersymmetric standard model (MSSM). The exact magnitude of the
bound depends on the scale of parity restoration and can be 10-20 GeV
above the MSSM bound if mass of the
right-handed $W_R$ is in the TeV range. An important
implication of our result is that since SUSYLR
models provide a simple realization of seesaw mechanism for neutrino
masses, measurement of the Higgs boson mass could provide an independent
probe of a low seesaw scale.
 \end{abstract}
\maketitle 

{\bf 1. Introduction} \ \ One of the main missing links of the otherwise immensely
successful Standard Model (SM) is the Higgs boson which plays the crucial role in giving
masses to all elementary particles in nature. It is therefore rightly the focus of a great deal of
theoretical~\cite{dawson} and experimental enquiries. Even though the Higgs boson mass in
SM is arbitrary, some ideas about how heavy the Higgs boson can be gained in the context
of different plausible extensions of SM as well as from other
considerations~\cite{quigg,sher}. Typical upper limits from, say unitarity
considerations~\cite{quigg} is in the TeV range. This bound is however considerably
strengthened in one of the most widely discussed possibility for TeV scale physics,
supersymmetry. Specifically in the minimal supersymmetric SM (MSSM), the upper bound on
the Higgs boson mass is $M^{max}_h\leq 135$ GeV~\cite{haber} when one and two loop
radiative corrections are included. Present collider searches provide a lower bound on
the SM Higgs mass~\cite{lep} of 114 GeV leaving a narrow region which need to be probed
to test MSSM. If the Higgs mass is found to be above this upper limit, does it mean that
supersymmetry is not relevant for physics at TeV scale? The answer is of course ``No''
since there exist simple and well motivated extensions of MSSM, e.g. the next-to-MSSM,
which extends the MSSM only by the addition of a singlet field~\cite{wyler} where there
is a relaxation of this bound to about 142 GeV or so~\cite{hnmssm}. There are also other
examples in literature\cite{babu} where simple modifications of the post-MSSM physics can
provide additional room for Higgs mass.

In this paper we discuss an alternative scenario motivated by neutrino mass as well as
understanding of the origin of parity violation~\cite{goran} where the upper limit on the
light Higgs mass is relaxed compared to MSSM. The model is the supersymmetric left-right
model (SUSYLR) \cite{susylr,kuchi} with TeV scale parity restoration (or TeV right-handed gauge boson mass $W_R$).
The change in the Higgs mass upper limit comes from the contribution of the D-terms and
satisfies the decoupling theorem i.e. as the $W_R$ mass goes to infinity, the Higgs mass
upper bound coincides with that for MSSM. This effect is to be expected on general
grounds~\cite{batra} in gauge extensions of MSSM.

\bigskip {\bf 2. Basics of the SUSYLR model} \ \  The gauge group of this
model is $SU(2)_L\times SU(2)_R\times U(1)_{B-L}\times SU(3)_c$.
The chiral left-handed and right-handed quark superfields are
denoted by $Q\equiv (u,d)$ and $Q^c\equiv (u^c, d^c)$ respectively
and similarly the lepton superfields are given by $L\equiv (\nu,
e)$ and $L^c\equiv (\nu^c, e^c)$. The $Q$ and $L$ transform as
left-handed doublets with the obvious values for the $B-L$ and the
$Q^c$ and $L^c$ transform as the right-handed doublets with
opposite $B-L$ values. The symmetry breaking is achieved by the
following set of Higgs superfields: $\Phi_a(2,2,0,1)$ ($a=1,2$);
$\Delta (3, 1, +2, 1)$; $\bar{\Delta}(3,1,-2,1)$;
$\Delta^c(1,3,-2,1)$ and $\bar{\Delta^c} (1,3,+2,1)$. We include a
gauge singlet superfield $S$ to facilitate the right handed
symmetry breaking. The symmetry breaking can also be carried out
by $B-L=1$ doublet fields, for which our results also apply. A
virtue of using triplet Higgs fields is that they lead to the
see-saw mechanism for small neutrino masses using only
renormalizable couplings. In addition, as was noted many years
ago~\cite{kuchi}, low scale $W_R$ requires that R-parity must
break spontaneously. This leads to many interesting
phenomenological implications that we do not address here.

The superpotential for the model is given by:
\begin{eqnarray}
 W &=& h_Q^a Q^T \tau_2 \Phi_a \tau_2 Q^c + h_L^a L^T \tau_2 \Phi_a
 \tau_2 L^c \nonumber \\
 &+& i f \left( L^T \tau_2 \Delta L + L^{cT} \tau_2 \Delta^c L^c
\right) + \mu_{ab} {\rm Tr} \left( \Phi_a^T \tau_2 \Phi_b \tau_2 \right) \nonumber \\
 &+& S
 \left[ {\rm Tr} \left( \Delta \bar \Delta + \Delta^c \bar \Delta^c
\right) - v_R^2 \right]
\end{eqnarray}
In order to analyze the Higgs mass spectrum, we write down the Higgs potential for the
model including the soft SUSY-breaking terms:
\begin{eqnarray}
V~=~V_F~+~V_S~+~V_D
\end{eqnarray}
where $V_F$ and $V_D$ are the standard F-term and D-term potential
and $V_S$ is the soft-SUSY-breaking terms which can be found in
the literature \cite{kuchi,huitu1}.
Minimization of the Higgs potential leads to the following vacuum
configuration for the $\Delta^c$ and $\nu^c$ Higgs
fields\cite{kuchi}:
\begin{eqnarray}
  \langle \widetilde L_i^c \rangle = \left( \begin{array}{c}
\langle \widetilde{\nu^c}\rangle \delta_{i1} \\
0
\end{array} \right), \ \langle\Delta^c \rangle = \left( \begin{array}{cc}
0 & 0 \\
\frac{v_R}{\sqrt{2}} & 0
\end{array} \right), \
 \langle\bar \Delta^c\rangle = \left( \begin{array}{cc}
0 & \frac{\bar v_R}{\sqrt{2}} \\
0 & 0
\end{array} \right) \nonumber
\end{eqnarray}
Note that in the SUSY limit $v_R=\bar v_R$ and $\langle
\widetilde{\nu^c}\rangle=0$. In the presence of supersymmetry
breaking terms however, $\langle \widetilde{\nu^c}\rangle$ is
nonzero.
On the other hand, if this model is extended to include a B-L=0
right handed triplet with nonzero vev, there appears a global
minimum of the potential which has $\langle
\widetilde{\nu^c}\rangle=0$~\cite{kuchi1} even in the presence of
susy breaking terms. Since the vev of $\nu^c$ is not relevant to
our discussion, we will work with B-L=0 triplet model and set
$\langle \widetilde{\nu^c}\rangle=0$ henceforth.
 The SM symmetry remains unbroken at this stage and is broken by
the vevs of the $\Phi$ fields. We can write these fields in terms
of their MSSM Higgs content:
\begin{eqnarray}
\Phi_i &=& \left( \begin{array}{cc}
\phi_{id}^0 & \phi_{iu}^+ \\
\phi_{id}^- & \phi_{iu}^0
\end{array}\right) \equiv \left( H_{di}, H_{ui} \right),
\end{eqnarray}
with vevs $ \langle H_{ui}^0 \rangle = \kappa_i, \  \langle H_{di}^0 \rangle =
\kappa'_i.$

Before  proceeding to discuss upper bound on the light neutral
Higgs mass in this model, we wish to make a few comments on the
implications of the TeV scale $W_R$ models for neutrino masses.
First, in the non-SUSY left-right model where neutrino mass has
both type I and type II seesaw contributions, having a TeV scale
$W_R$ is unnatural since the type II seesaw contribution then
becomes extremely large. This is due to the presence of non-zero
couplings of type ${\rm Tr}
\phi\Delta_L\phi^\dagger\Delta^\dagger_R$, which are allowed by
the symmetries of the theory. On the other hand, in the SUSYLR
model, this coupling is absent due to supersymmetry and therefore
there is no type II seesaw contribution to neutrino mass. As far
as the type I contribution is concerned, if we choose $h_\nu\sim
h_e$ where $h_\nu$ is the Dirac neutrino Yukawa coupling, then we
can have a few TeV $W_R$ and neutrino masses of order of eV. Thus
as far as neutrino masses go, low scale $W_R$ is a realistic
model.

\bigskip

{\bf 3. Light Higgs mass bound: single bi-doublet case}  \ \ We proceed to consider the
bound on the light neutral Higgs mass in the SUSYLR model. We work in the limit where
$v_R$ and $\bar v_R$ are much bigger than the SM scale. In this limit, we search for
additional contributions to the MSSM Higgs potential, which will be at the heart of the
change in the upper limit of the Higgs boson mass.

We first illustrate this in a one bi-doublet model. This simple model leads to vanishing
CKM angles at the tree level, which can be fixed in one of two ways: (i) by including
radiative correction effects from squark mixings~\cite{dutta} or (ii) by including a
second bi-doublet which decouples from the low energy sector but it has a tadpole induced
vev that generates the correct CKM angles. We discuss the case (ii) toward the end of the
paper. The interesting point is that neither of these affects the Higgs mass upper bound
that we derive.
For the model under consideration, we first show that in the SUSY limit the low energy
Higgs potential recovers that of MSSM in the same limit. When soft SUSY breaking terms is taken into account, there appear new contributions to the MSSM Higgs potential, serving to raise the upper limit on the light Higgs mass, which can be significant for TeV scale $W_R$.

We start with a review of the well known symmetry breaking of the model by the triplet
Higgs fields   $\Delta^c$ and $\bar \Delta^c$ in the SUSY case. The gauge bosons get mass
from the kinetic terms of triplets and after symmetry breaking, the massless gauge boson
and gaugino corresponding to $U(1)_Y$ is the combination $B = \frac{g_{BL}}{\sqrt{g_R^2 +
g_{BL}^2}} W_{3R} +
 \frac{g_R}{\sqrt{g_R^2 + g_{BL}^2}} V_{BL}$, with the hypercharge gauge
coupling given by $ \displaystyle\frac{1}{g^2_Y}~=~\frac{1}{g^2_R}+\frac{1}{g^2_{BL}}$.
The heavy $Z'$-boson has mass squared $M_{Z'}^2 = 2 (g_R^2 + g_{BL}^2) v_R^2$. There is a
factor 2 compared with the charged $W_R$ boson mass $M_{W_R^\pm}^2 = g_R^2 v_R^2$
because the triplet vev breaks custodial symmetry for the right-handed sector.

Since there is no coupling between the bidoublet Higgs and the triplet Higgs fields
responsible for parity breaking in Eq. (1), any change in the effective MSSM doublet
Higgs potential below the $v_R$ scale must originate from the D-terms,
\begin{eqnarray}
V_D &=& \frac{g_R^2}{8} \left| {\rm Tr} [ 2 \Delta^{c\dag} \tau_m \Delta^c
+ 2 \bar \Delta^{c\dag} \tau_m \bar \Delta^c + \Phi \tau_m^T \Phi^\dag ]
\right|^2  \nonumber \\
&+&  \frac{g_{BL}^2}{8} \left( {\rm Tr} [ 2 \Delta^{c\dag} \Delta^c - 2 \bar
\Delta^{c\dag} \bar \Delta^c ] \right)^2 \ .
\end{eqnarray}
The contribution to the neutral Higgs fields couplings is,
\begin{eqnarray}\label{V}
V_D^{\rm neut.} &=& \frac{g_R^2}{8}  \left| {\rm Tr} [ \Phi \tau_3^T \Phi^\dag ]
\right|^2 + 4 (g_R^2 +  g_{BL}^2) v_R^2 \left| \frac{\Delta^{c0} - \bar
\Delta^{c0}}{\sqrt{2}}\right|^2 \nonumber \\
&+& g_R^2 v_R {\rm Re}[\left(\Delta^{c0} - \bar \Delta^{c0}\right)] {\rm Tr} [ \Phi
\tau_3^T \Phi^\dag ] \ .
\end{eqnarray}
The coupling is linear in the field Re$[\left( \Delta^{c0} - \bar\Delta^{c0}\right)]$,
which we will call $\sigma_-$. As $\sigma_-$ field becomes heavy, its coupling to $[\Phi
\tau_3^T \Phi^\dag ]$ will generate new quartic term in the MSSM doublet field potential,
which in turn will lead to new contributions to Higgs mass upper bound. Collecting this
new effect, we get for the Higgs quartic term:
\begin{eqnarray}\label{quartic}
\delta V(\Phi)~= ~\frac{1}{8} \left( g_R^2 -
\frac{g^4_Rv^2_R}{M^2_{\sigma_-}}\right) \left| {\rm Tr} [ \Phi
\tau_3^T \Phi^\dag ] \right|^2
\end{eqnarray}
To evaluate this new contribution, we need to know $M_{\sigma_-}$. This has two potential
contributions: (i) from the D-term and (ii) from the F-term contribution to the Higgs
potential. It turns out that in the SUSY limit, the only contribution to $M_{\sigma_-}$
is from the D-terms and we have $M^2_{\sigma_-}=2(g^2_R+g^2_{BL})v^2_R$. This follows not
only from actual calculations but also from the fact that $\sigma_-$ is a member of the
Goldstone supermultiplet, all members of which must have the same mass as $Z'$ in the
SUSY limit. This result would hold even if the superpotential had a term of the
form $\mu \Delta^c\bar{\Delta^c}$. Using this in Eq. (\ref{quartic}), it is easy to see
that the net contribution to the quartic term in the Higgs superpotential becomes
$\displaystyle\frac{g^2_Y}{8}\left(|H_u|^2-|H_d|^2\right)^2$. This is nothing but the
$D_Y$ contribution to quartic Higgs doublet term in MSSM. Since in the decoupling limit,
we get MSSM, as expected from the decoupling theorem.

Let us now switch on the supersymmetry breaking terms. In their
presence, the $\sigma_-$ field has  aditional contributions which
lead to a shift in the Higgs masses. To see this we introduce soft
mass term $m_S^2 S^\dag S$ as well as SUSY breaking mass terms for
the $\Delta^c$ and $\bar{\Delta^c}$. Taking the same SUSY breaking
terms for all the fields gives different value for the $\sigma_-$
field mass and we get for the contribution to the quartic Higgs term
$\displaystyle\frac{g^2_{Y,eff}}{8}\left(|H_u|^2-|H_d|^2\right)^2$
where
\begin{eqnarray}\label{xxx}
g^{2}_{\rm Y, eff} = g_R^2 - \frac{g_R^4}{g_R^2 + g_{BL}^2 + \frac{m_0^2}{2 v_R^2}} =
\frac{g_R^2 g_{BL}^2 + g_R^2  \frac{m_0^2}{2v_R^2}}{g_R^2 + g_{BL}^2 +
\frac{m_0^2}{2v_R^2}}
\end{eqnarray}
where $m_0$ in the above equation is a generic soft mass term for
sparticles that breaks supersymmetry. This leads to an enhancement
of the Higgs mass upper bound since $g^{2}_{\rm Y, eff} >
g^{2}_{\rm Y, SM}$ . To get an idea about how large the change in
the upper bound is likely to be, we take $m_0 = 1$ TeV, $v_R = 2$
TeV, then from $g_L^2 \approx 0.42$ and $g_Y^{2} \approx 0.13$, we
get the ratio $ r = \frac{g_L^2 + g^{2}_{\rm Y, eff}}{g_L^2 +
g^{2}_Y} \approx 1.1$, which will give 10\% increase of the tree
level upper bound on the light Higgs mass i.e. it increases from
$M_Z = 90$ GeV to 100 GeV.

It is also worth pointing out that as the scale of parity violation goes to infinity,
this new contribution goes to zero and one recovers the MSSM result. This is an important
consistency check on our result~\cite{pandita}.

\bigskip {\bf 4. One-loop radiative corrections and numerical result} \ \ In the following, we
discuss the radiative corrections to the Higgs boson mass for the model above. It is well
known that
\begin{eqnarray}\label{standard}
 \Delta V_1 = \frac{1}{64 \pi^2} {\rm Str}\ \mathcal{M}^4 \left( \log
\frac{\mathcal{M}^2}{Q^2} - \frac{3}{2} \right)\ ,
\end{eqnarray}
where the supertrace ${\rm Str}$ means $ {\rm Str}\ f(\mathcal{M}^2) = \sum_i (-1)^{2
J_i} (2 J_i + 1) f(m_i^2)$, where the mass $m_i$ is calculated in background fields, and
the sum counts all the fermionic and bosonic degrees of freedom. $J_i$ is the spin for
particle $i$ and $Q$ is the renormalization scale on the order of electroweak symmetry breaking. In MSSM, top
and stop contributes dominantly to $\Delta V_1$ because of large Yukawa coupling $y_t$,
while the bottom and sbottom contribution is only important for $\tan \beta \gg 20$.

 In SUSYLR model with one bidoublet, we have $\tan\beta = m_t/m_b \approx 40$,
 so the sbottom quark can couples to $H_u$ with a large coupling $y_t$. This
is from the F-term of bidoublet field $\Phi_1$.
\begin{eqnarray}
 (F_{\Phi_1})_{ij} &=& h_1 (Q_L^T \tau_2)_i (\tau_2 Q_R^c)_j + \mu_{11}
 (\tau_2 \Phi_1^T \tau_2)_{ji}
 \\
 |F_{\Phi_1}|^2 &=& \mu_{11} m_t \cot \beta \cdot \widetilde
 t_L^{\dag} \widetilde t_R + \mu_{11} m_t \cdot \widetilde b_L^{\dag}
\widetilde b_R + \cdots \nonumber
\end{eqnarray}
Note the second term in the second line differs from MSSM by a factor $m_t / (m_b \tan\beta)$, since $H_u$ and
$H_d$ are unified into the same bidoublet. This means the sbottom mass receives a large
LR mixing proportional to the top Yukawa couplng, even for low $\tan\beta$ (in the
presence of a second bidoublet in the realistic model below). Therefore now there are
 three fields that have to be taken into account: top, stop and sbottom.
 If we neglect the small couplings except for $y_t$, and also neglect the
$A$-terms
their masses can be approximated by
\begin{eqnarray}
m_t^2 &=& y_t^2 |H_u|^2 \nonumber \\
 m_{\widetilde t_1}^2 &\simeq& m_{\widetilde t_2}^2 \simeq m_{\widetilde
Q}^2 + y_t^2 |H_u|^2 \nonumber \\
 m_{\widetilde b_1}^2 &\simeq& m_{\widetilde Q}^2 + y_t \mu_{11} |H_u| +
\cdots \nonumber \\
 m_{\widetilde b_2}^2 &\simeq& m_{\widetilde Q}^2 - y_t \mu_{11} |H_u| +
\cdots
\end{eqnarray}
The $\cdots$ represents dependence
 on $|H_d|$, but since the lightest Higgs boson is mainly made up of
 $H_u$ for large $\tan\beta$ with only a small $H_d$ component, this
dependence can be neglected.
We can choose a proper scale $Q$ so that the first derivative vanishes,
which will be
important in eliminating the explicit $Q$
 dependence of Higgs mass, i.e. it can only depend on $Q$ through
depending on other parameters.
Second derivative gives radiative corrections to the lightest Higgs mass
\begin{eqnarray}\label{111}
\delta M_{h}^2 &\simeq& \frac{1}{2}\frac{\partial^2 \Delta V_1}{\partial
|H_u|^2} =
 \frac{3g_L^2}{8 \pi^2} \frac{m_t^4}{M_{W_L}^2} \log \frac{m_{\widetilde
 t}^2}{m_t^2} \nonumber \\
 &-& \frac{3 g_L^2}{64 \pi^2} \frac{m_0^2 \mu_{11} m_t}{M_{W_L}^2}
 \log \frac{m_{\widetilde b_1}^2}{m_{\widetilde b_2}^2} +
\frac{3g_L^2}{32\pi^2} \frac{\mu_{11}^2 m_t^2}{M_{W_L}^2}
\end{eqnarray}
 The first term is as the usual MSSM one. Now we have two new terms
 proportional to $\mu_{11}$. Their sum is an even function of $\mu_{11}$
since changing the sign also interchanges $\widetilde b_1 \leftrightarrow
 \widetilde b_2$. We find the sum of second and third term are negative
definite for arbitrary $\mu_{11}$.
 Actually, for $m_0 \sim 1$ TeV, one can expand with
 $\frac{\mu_{11}m_t}{m_0^2}$. The net contribution is non-vanishing only
 up to the third order, which is $-\frac{g_L^2}{32 \pi^2}
 \frac{\mu_{11}^4 m_t^4}{M_{W_L}^2 m_0^4}$, depending on $\mu_{11}$ very
 mildly. For $\mu_{11}$ ranging from 100 GeV to 300 GeV around EW scale,
 this negative contribution is smaller than 1 GeV. Note that in this
case, we did not have to discuss the details of
EWSB since it is very similar to MSSM.

 Let us present the numerical results for the Higgs mass upper bound
for this
 scenario. In Fig.~1, we plot the difference in the prediction of upper
bound on the
 lightest Higgs boson mass between SUSYLR model and MSSM: $\Delta M_h = m_h^{\rm SUSYLR}-m_h^{\rm MSSM}$. For the right-handed scale near 2-3 TeV, the upward shift of the Higgs mass
bound can be of a few GeV, increasing as $v_R$ decreases.
(For symmetry breaking using Higgs doublets, $M_{\sigma^-}^2 = (g_R^2+g_{BL}^2)v_R^2$ and $g_{\rm Y, eff}$ in Eq.~(\ref{xxx}) gets increased to $\frac{g_R^2 g_{BL}^2 + g_R^2 m_0^2/v_R^2}{g_R^2 + g_{BL}^2 + m_0^2/v_R^2}$. Then $\Delta M_h$ can further increase by a factor of 2.)
The bound also increases with the soft mass scale
$m_0$ as expected. From the discussions below Eq.~(\ref{111}),
non-zero $\mu_{11}$ always gives small and negative contribution. In order not to violate the lower bound on
chargino mass at LEP2, we choose $\mu_{11}
\geq 100$ GeV.


\begin{figure}[hbt]
\begin{center}
\includegraphics[width=7cm]{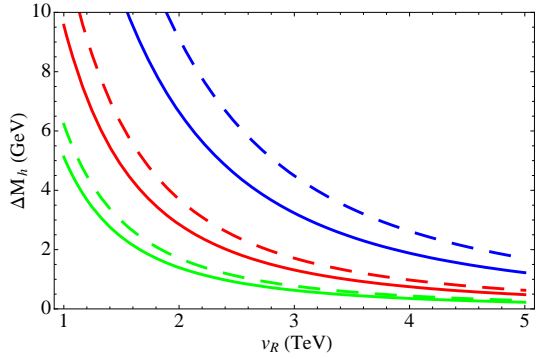}
\caption{Higgs mass bound from SUSYLR model shown as the difference from MSSM. Tree level
results are shown as dashed curves and radiative corrected ones are solid curves. Upper,
middle and lower two curves correspond to $m_0=1$ TeV (blue), $m_0=600$ GeV (red),
$m_0=400$ GeV (green), respectively. For radiative correction, we choose $\mu_{11} = 100$
GeV. For all the curves, we choose $\tan \beta \approx 40$ as noted.}
\end{center}
\end{figure}


\bigskip
{\bf 5. Extension to two bi-doublet case} \ \ We can extend our
discussion to the more realistic two bi-doublet case as needed to
generate the correct CKM angles at the tree level. There are two
possible ways to do that; both these we discuss below.

\noindent{\bf Model A}: 

In this case, we identify $\Phi_1$ as the
bi-doublet of the previous section. We can diagonalize the
corresponding Yukawa coupling matrix $h^1_Q$. We are then forced
to have all elements of the second bi-doublet $\Phi_2$ Yukawa
coupling, $h^2_Q\neq 0$. Once the second $\Phi_2$ has vev, by
appropriate choice of this matrix, we can generate the desired
quark masses and mixings. We will see that even though there are
four real neutral Higgs fields in this case, the upper limit on
the light Higgs field remains the same as in the single bi-doublet
case.

To see this, first we choose a basis in the $\Phi$ space such that the
superpotential for $\Phi$'s has the form
\begin{eqnarray}
W_{extra}~=~\mu_{11} {\rm Tr}(\Phi_1\Phi_1)~+~\mu_{22} {\rm Tr}(\Phi_2\Phi_2)
\end{eqnarray}
We assume that $\mu_{22} \gg \mu_{11}$. We have then no freedom to
diagonalize the soft SUSY breaking terms. The sum of the $V_F+V_S$ can
then in general be written as
\begin{eqnarray}
V_F~+~V_S(\Phi_i)&=&m^2_{ij}{\rm Tr}\phi^\dagger_i\phi_j~+~b_{11}
{\rm Tr}(\phi_1\phi_1) \nonumber \\
&+&b_{22} {\rm Tr}(\phi_2\phi_2)+~{\rm h.c.}
\end{eqnarray}
Note that if $m^2_{22} \gg m^2_{12}$, then the mixed term in the
$\phi$'s will induce a vev for the $\phi_2$ field which is small
compared to that for the $\phi_1$ i.e. $\kappa_2, \kappa'_2 \ll
\kappa_1,\kappa'_1$. To generate the correct mass and mixing
pattern for the quarks, it is sufficient to have the $\phi_2$ vevs
of order of a 100 MeV. For instance if $m_{12}\leq 10 $ GeV and
$m_{22} = 1$ TeV, then we can estimate $\kappa'_2 = \kappa'_1
m_{12} / m_{22} = \kappa_1'/100 \sim 100$ MeV, which is enough to
generate the strange quark mass as well as other CKM angles. Note
also that one should include the one loop effects coming from
squark masses and mixings\cite{dutta}. While we do not give a
detailed fit here, it seems clear that this is a realistic model
where the new Higgs mixing parameter $m_{12}$ is in the 1-10 GeV
range. When it is close to one GeV, the effect on the Higgs mass
upper bound is also about a GeV lower due to off diagonal
contributions. We can also keep the $\Phi_2$ Yukawa couplings
sufficiently small so that their radiative corrections do not
affect the one loop result. This vacuum then is a perturbation
around the vacuum of the single bi-doublet case and furthermore
due to large $\mu_{22}$, the $H_{u,d}$ coming from the second
bi-doublet will acquire heavy mass and decouple without affecting
the light Higgs mass upper bound except perhaps a small one GeV or
so shift. This case corresponds to large tan$\beta \approx$40.

\noindent{\bf Model B:}

In this case, we choose two bi-doublets with the vev pattern given
by:
\begin{eqnarray}
\langle\Phi_1 \rangle &=& \left( \begin{array}{cc}
\kappa & 0\\
0 & 0
\end{array}\right), \ \ \ 
\langle\Phi_2 \rangle = \left( \begin{array}{cc} 0 & 0 \\0 & \kappa'
\end{array}\right)
\end{eqnarray}
Since the down quark masses in this case come from a second Yukawa 
coupling unlike the model A, we can have the 
value of tan$\beta ~\equiv\frac{\kappa'}{\kappa}$ much lower than 40 by 
appropriate choice of the second Yukawa coupling matrix. There are 
generally four electreweak scale Higgs doublets. Using the 
standard formula in Eq.~(\ref{standard}), 
we have calculated the 1-loop radiative corrections to the $4 \times 4$ 
neutral Higgs mass matrix in the presence of SUSY breaking thresholds.
As before, we negleted all other couplings but $y_t$ when calculating 
$\Delta V_1$. We also keep the effect of non-zero
vev for the right-handed sneutrino. However due to small neutrino Dirac 
Yukawa couplings, the mixing effect between the Higgs field and the 
left-handed sneutrino caused by the right-handed sneutrino 
vev is very small and does not affect our result. In order to estimate 
the upper bound, we have done a numerical study to obtain the Higgs mass 
for random choice of parameters. The results are the scatter points in 
Fig. 2 below for a 
choice of the generic soft mass scale $m_0=1$ TeV and right-handed scale 
$v_R=1.5$ TeV. Each point in the 
scatter plot represents the lightest Higgs mass for a specific choice of 
parameters. The upper limit therefore corresponds to the topmost set of 
points in Fig. 2. In contrast, the MSSM Higgs mass 
upper bound is plotted as the yellow (lower) curve, which is at most 130 GeV after 
1-loop radiative corrections with the same choice of $m_0$. 
The red (upper) curve is for Model A. We find Fig. 2 that in general SUSYLR model 
the upper bound can be as high as 140 GeV or even more especially 
in the regime $5<\tan\beta<10$. This is higher than the prediction of 
MSSM. Clearly, as the right-handed scale goes down, the upper bound 
inceases.

\begin{figure}[hbt]
\begin{center}
\includegraphics[width=7cm]{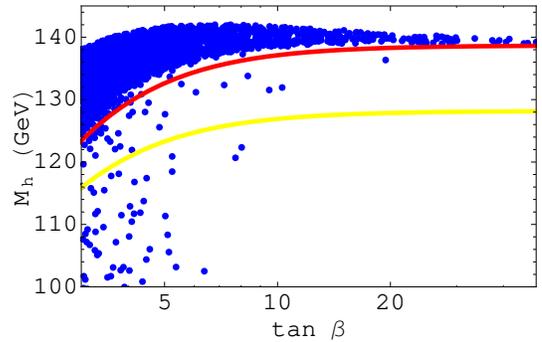}
\caption{The scatter plot represents the Higgs mass values as 
a function of tan$\beta$ in
the two bi-doublet SUSYLR model for random choice of parameters.
The yellow (lower) curve is the prediction for the light Higgs mass upper bound 
in MSSM, while the red (upper) curve is for 
Model A in SUSYLR. The blue points are for Model B in the SUSYLR case. 
When plotting the figure, we choose $m_0=1$ TeV and $v_R=1.5$ TeV.
We have used a Monte Carlo simulation for the parameter space to 
generate the plot for Model B. }
\end{center}
\end{figure}

{\bf 6. Comments and Conclusion:} Before concluding, we wish to
make a few comments on the model:

(i) Low scale non-SUSY left-right models have strong constraints
coming from the tree level Higgs contribution to flavor changing
processes. In the SUSY version however, there are additional
contributions to the same from squark and slepton sector which can
be used to cancel this effect\cite{ji1}. While strictly this is
not natural, from a phenomenological point of view, this makes the
model consistent when both the $W_R$ and Higgs masses are in the
few TeV range.

(ii) The second point is that unlike other models such as NMSSM
where the light Higgs mass bound is changed by making additional
assumptions about the Higgs couplings (e.g. not hitting the Landau
pole at the GUT scale), in our model the increase in the bound is
purely gauge coupling induced and is independent of the Higgs
couplings.

(iii) It is also worth stressing the obvious point that
observation of a Higgs with mass above the MSSM bound of 135 GeV
is not necessarily an evidence for the SUSYLR model since there
exist other models with which also relax this bound. One needs
other direct evidences such as the mass of $W_R$ or $Z'$ produced
at LHC which when combined with observed higher Higgs mass could
provide evidence for low left-right seesaw scale.

 To conclude, we have pointed
out that the upper bound on the light Higgs mass is higher if MSSM
is assumed to be an effective low energy theory of a TeV scale
SUSYLR model. The increase can be as much as 10 GeV or more
depending on the scale of parity breaking. If the Higgs boson mass
in the collider searches is found to exceed the MSSM upper limit
of 135 GeV, one interpretation of that could be in terms of a TeV
scale seesaw in the context of a SUSYLR model.

 We thank K. Agashe, P. Batra, and Z. Chacko for comments. This work was 
partially supported by the U. S. Department of Energy via grant 
DE-FG02-93ER-40762. R. N. M. is supported by NSF grant No. PHY-0652363. 
Y. Z. acknowledges the hospitality and support
 from the TQHN group at University of Maryland and a partial support from 
NSFC grants
10421503 and 10625521.




\begin{thebibliography}{90}


\bibitem{dawson} For a review, see S. Dawson, J. Gunion, H. Haber and G.
Kane, {\it Higgs Hunter's guide}, SCIPP-89/13, UCD-89-4,
BNL-41644, Jun 1989. 404pp.


\bibitem{quigg} for an upper limit from unitarity considerations, see
 B.~W.~Lee, C.~Quigg and H.~B.~Thacker, Phys.\ Rev.\  D {\bf 16}, 1519 (1977).

\bibitem{sher} M.~Sher, Phys.\ Rept.\  {\bf 179}, 273 (1989).


\bibitem{haber} H. Haber and R. Hemfling, Phys. Rev. Lett. {\bf 66}, 1815
(1991); J.~R.~Ellis, G.~Ridolfi and F.~Zwirner,
  Phys.\ Lett.\  B {\bf 257}, 83 (1991).

\bibitem{lep} R. Barate et al. Phys. Lett. {\bf 565}, 61 (2003).

\bibitem{wyler} H.~P.~Nilles, M.~Srednicki and D.~Wyler,
  Phys.\ Lett.\  B {\bf 120}, 346 (1983).


\bibitem{hnmssm} U.~Ellwanger and C.~Hugonie,  Mod.\ Phys.\ Lett.\  A
{\bf 22}, 1581 (2007).

\bibitem{babu} Some examples of such models with relaxed Higgs mass limit are:
M.~Drees, Int.\ J.\ Mod.\ Phys.\  A {\bf 4}, 3635 (1989);
  K.~S.~Babu, I.~Gogoladze and C.~Kolda, arXiv:hep-ph/0410085; G.~Bhattacharyya, S.~K.~Majee and A.~Raychaudhuri,
  Nucl.\ Phys.\  B {\bf 793}, 114 (2008); S.~Hesselbach, D.~J.~Miller, G.~Moortgat-Pick, R.~Nevzorov and M.~Trusov,
  Phys.\ Lett.\  B {\bf 662}, 199 (2008); M.~Dine, N.~Seiberg and S.~Thomas,
  Phys.\ Rev.\  D {\bf 76}, 095004 (2007); V.~Barger, P.~Langacker, H.~S.~Lee and G.~Shaughnessy,
  Phys.\ Rev.\  D {\bf 73}, 115010 (2006).


\bibitem{goran} R. N. Mohapatra and J. C. Pati, Phys. Rev. {\bf D 11},
566, 2558  (1975); G. Senjanovi\'c and R. N. Mohapatra, Phys. Rev. {\bf D
12}, 1502 (1975).

\bibitem{susylr} M.~Cvetic and J.~C.~Pati, Phys.\ Lett.\  B {\bf 135},
57 (1984); R.~M.~Francis, M.~Frank and C.~S.~Kalman,
  Phys.\ Rev.\  D {\bf 43}, 2369 (1991).

\bibitem{kuchi} R.~Kuchimanchi and R.~N.~Mohapatra, Phys.\ Rev.\  D {\bf
48}, 4352 (1993).

\bibitem{huitu1}  K.~Huitu, P.~N.~Pandita and K.~Puolamaki,
  arXiv:hep-ph/9904388.


\bibitem{batra} P.~Batra, A.~Delgado, D.~E.~Kaplan and T.~M.~P.~Tait,
  JHEP {\bf 0402}, 043 (2004).

\bibitem{kuchi1} R.~Kuchimanchi and R.~N.~Mohapatra,
  Phys.\ Rev.\ Lett.\  {\bf 75}, 3989 (1995)

\bibitem{dutta}  K.~S.~Babu, B.~Dutta and R.~N.~Mohapatra,
  Phys.\ Rev.\  D {\bf 60}, 095004 (1999).


\bibitem{pandita} The upper bound on light Higgs mass in SUSYLR models
derived in K.~Huitu, P.~N.~Pandita and K.~Puolamaki,
  Phys.\ Lett.\  B {\bf 423}, 97 (1998) does not satisfy this decoupling
consistency.


\bibitem{ji1} 
  Y.~Zhang, H.~An and X.~d.~Ji,
  arXiv:0710.1454 [hep-ph].


\end{thebibliography}
\end{document}